\documentclass{article}
\usepackage{eurosym}
\usepackage{amssymb}
\usepackage{amsmath}
\usepackage{cite}

\begin{document}

\title{Symmetry Analysis for a Fourth-order Noise-reduction Partial
Differential Equation}
\author{Andronikos Paliathanasis\thanks{%
Email: anpaliat@phys.uoa.gr} \\
{\ \textit{Institute of Systems Science, Durban University of Technology }}\\
{\ \textit{PO Box 1334, Durban 4000, Republic of South Africa}} \\
{\ \textit{Instituto de Ciencias F\'{\i}sicas y Matem\'{a}ticas,}}\\
{\ \textit{Universidad Austral de
Chile, Valdivia, Chile}}
\and P.G.L. Leach \\
{\ \textit{Institute of Systems Science, Durban University of Technology }}\\
{\ \textit{PO Box 1334, Durban 4000, Republic of South Africa}} \\
{\ \textit{School of Mathematical Sciences, University of KwaZulu-Natal }}\\
{\ \textit{Durban, Republic of South Africa}}}
\maketitle

\begin{abstract}
We apply the theory of Lie symmetries in order to study a fourth-order $1+2$
evolutionary partial differential equation which has been proposed for the
image processing noise reduction. In particular we determine the Lie point
symmetries for the specific 1+2 partial differential equations and we apply
the invariant functions to determine similarity solutions. For the static
solutions we observe that the reduced fourth-order ordinary differential
equations are reduced to second-order ordinary differential equations which
are maximally symmetric. Finally, nonstatic closed-form solutions are also
determined.

\end{abstract}

\section{Introduction}

During the last three decades significant work has been done on the subject
of noise reduction in image processing with the use of partial differential
equations (PDEs). By noise reduction or noise removal we mean the algorithm
which has to be applied to a digital image to remove the wrong data
information, or artificial imprints added to the digital image by the
sensors; while during this procedure it is important that the original
information and structure to be preserved \cite{marr}. \ For a review on the
various technics on noise reduction methods we refer the reader in \cite%
{rev1}.

The linear heat equation was one of the first PDEs to which was applied
image reduction \cite{catte,wei1}, while the introduction of nonlinear terms
into the PDE \cite{ref1} was necessary in order to avoid the flatness of a
noise image. Some other nonlinear models can be found and are presented for
instance in \cite{ref2,ref3,ref4,ref5}.

You and Kaveh in \cite{you} proposed the following nonlinear higher-order
diffusion equation\footnote{$\Delta $ denotes the Laplace operator.}%
\begin{equation}
\frac{\partial u}{\partial t}=-E_{L}\left( f\left\vert \Delta u\right\vert
\right) ,  \label{eq.01}
\end{equation}%
where the right hand side of (\ref{eq.01}) follows from the variational
principle of the Action Integral, $S=\int f\left( \left\vert \Delta
u\right\vert \right) d\mathbf{x~}$. In the case of a two-dimensional flat
space this becomes%
\begin{equation}
S=\int f\left( \left\vert u_{,xx}+u_{,yy}\right\vert \right) dxdy.
\label{eq.02}
\end{equation}%
Consequently, $f$ is the Lagrange function, while a second requirement for
function $f$ is $f^{\prime }>0.$

Now, in\ the simplest case for which $f$ is a linear function, i.e. $f\left(
\left\vert \Delta u\right\vert \right) =\sqrt{u_{,xx}+u_{,yy}}$, equation (%
\ref{eq.01}) takes the following form%
\begin{equation}
\frac{\left( u_{xxxx}+2u_{xxyy}+u_{yyyy}\right) \left( u_{xx}+u_{yy}\right)
-3\left[ \left( u_{xx}+u_{yy}\right) _{,x}\right] ^{2}-3\left[ \left(
u_{xx}+u_{yy}\right) _{,y}\right] ^{2}}{\sqrt{\left\vert
u_{,xx}+u_{,yy}\right\vert }}+u_{t}=0.  \label{eq.03}
\end{equation}%
That particular linear function, $f$, can be seen as a natural extension of
the Lagrangian for the second-order differential equation studied in \cite%
{ref1}. While equation (\ref{eq.03}) is a fourth-order equation, there are
similarities with the second-order minimal surface equation. In particular,
in the minimal surface equation the\ Action Integral involves the
minimization of a surface and not of an arc length, as is the case with the
equations of motion \cite{Bila, Peterson,Aqadir}. In this work, the equation
(\ref{eq.03}) follows from the minimization of a length which is defined by
second-order derivatives.

In this work we study the latter nonlinear fourth-order PDE by applying
Lie's theory on the symmetries of differential equations. Lie's theory is
one of the main mathematical tools for the determination of solutions for
nonlinear differential equations. The existence of a Lie symmetry for a
given differential equation is important because invariant functions can be
constructed to reduce the order of the differential equation or the number
of the dependent variables \cite{bluman}. Furthermore, Lie symmetries can be
used to perform group classifications for differential equations \cite%
{ovsiannikov} or identify well-known systems for instance see \cite%
{leach1,leach2,erm,cl1,cl2,cl3,cl4,cl5,cl6,cl7} and references therein. The
applications of Lie symmetries cover various areas of Mathematical physics
from liquid films and travel wave solutions \cite{cl8,cl9,cl10,cl11}, the
theory of diffusion \cite{cl12,cl13} and optical physics \cite{cl14}

The application of Lie symmetries to the theory of noise reduction is not
new. Recently in \cite{kara1} a group classification analysis was performed
for the nonlinear second-order PDE proposed by Rudin et al \cite{ref1} and
Lie symmetries were applied for the determination of conservation laws. In
this work we approach the problem differently, by applying the Lie
invariants to perform the reduction process for the PDE (\ref{eq.03}). The
importance of the application of Lie point symmetries on noise reduction
PDEs is not only that closed-form solutions can be derived, but Lie
symmetries can be applied also to recognize patterns in an image as proposed
by Big\"{u}n \cite{bigun,bigun2}.

The plan of the paper is as follows.

In Section \ref{sec2} the basic properties and definitions of Lie point
symmetries of differential equations are presented. Moreover, we present the
Lie symmetries for equation (\ref{eq.03}) and that of the time-independent
case. The application of the Lie invariants is performed in Section \ref%
{sec3} from where we find that the static solutions reduce always to
maximally symmetric second-order differential equations. That is not true
for the nonstatic invariant solutions. However, for the latter case we are
able to find some particular exact solutions. Finally we discuss our results
and draw our conclusions in Section \ref{sec4}.

\section{Lie point symmetries}

\label{sec2}

Let function $\Phi $ describe the map of an one-parameter point
transformation such as $\Phi \left( u\left( x^{i}\right) \right) =u\left(
x^{i}\right) \ $with infinitesimal transformation%
\begin{eqnarray}
t^{\prime } &=&t^{i}+\varepsilon \xi \left( t,x^{i},u\right)  \label{sv.12}
\\
x^{i\prime } &=&x^{i}+\varepsilon \xi ^{i}\left( t,x^{i},u\right) \\
u^{\prime } &=&u+\varepsilon \eta \left( t,x^{i},u\right)  \label{sv.13}
\end{eqnarray}%
and generator
\begin{equation}
\Gamma =\frac{\partial t^{\prime }}{\partial \varepsilon }\partial _{t}+%
\frac{\partial x^{\prime }}{\partial \varepsilon }\partial _{x}+\frac{%
\partial u}{\partial \varepsilon }\partial _{u},  \label{sv.16}
\end{equation}%
where~{$\varepsilon $ is the parameter of smallness} and $x^{i}=\left(
x,y\right) $. \

Let $u\left( x^{i}\right) $ be a solution of the PDE
\begin{equation}
\mathcal{H}\left( u,u_{,t},u_{,x}...\right) =0;
\end{equation}
then under the map $\Phi $, function $u^{\prime }\left( x^{i\prime }\right)
=\Phi \left( u\left( x^{i}\right) \right) $ is a solution for the latter
differential equation if and only if the differential equation is also
invariant under the action of the map, $\Phi $, i.e.
\begin{equation}
\Phi \left( \mathcal{H}\left( u,u_{,t},u_{,x}...\right) \right) =0.
\end{equation}

When the latter expression is true, the generator $\Gamma $ is called a Lie
point symmetry for the differential equation. It is straightforward to prove
that this condition becomes
\begin{equation}
\Gamma ^{\left[ n\right] }\left( \mathcal{H}\right) =0,  \label{sv.17}
\end{equation}%
in which $\Gamma ^{\left[ n\right] }$ describes the $n^{th}$%
prolongation/extension of the symmetry vector in the jet-space of variables,
$\left\{ t,x^{i},u,u_{,i},u_{,ij},...\right\} $.

The importance of the existence of a Lie symmetry for a given PDE is that
from the associated Lagrange's system,%
\begin{equation}
\frac{dt}{\xi ^{t}}=\frac{dx^{i}}{\xi ^{i}}=\frac{du}{\eta },
\end{equation}%
zeroth-order invariants,~$U^{\left[ 0\right] }\left( t,x^{i},u\right) $ are
able to be determined which can be used to reduce the number of the
independent variables of the differential equation and lead to the
construction of similarity solutions. Last but not least, the admitted
symmetries of a given differential equation constitute a Lie algebra.

\subsection{Lie point symmetries for the time-independent equation}

Consider now $u\left( t,x,y\right) =u\left( x,y\right) $. Then equation (\ref%
{eq.03}) is simplified as follows.

The resulting differential equation is of fourth-order and it is given as%
\begin{equation}
\left( u_{xxxx}+2u_{xxyy}+u_{yyyy}\right) \left( u_{xx}+u_{yy}\right) -3
\left[ \left( u_{xx}+u_{yy}\right) _{,x}\right] ^{2}-3\left[ \left(
u_{xx}+u_{yy}\right) _{,y}\right] ^{2}=0.  \label{eq.04}
\end{equation}%
For this time-independent equation the application of Lie's theories
provides the following $5+\infty $ Lie point symmetries.
\begin{eqnarray*}
X_{1} &=&\partial _{x}~~,~X_{2}=\partial _{y}~~,~~X_{3}=y\partial
_{x}-x\partial _{x}~,~X_{4}=u\partial _{u} \\
X_{5} &=&x\partial _{x}+y\partial _{y}~,~X_{\infty }=\Psi \left( x,y\right)
\partial _{u}~\text{with }\Delta \Psi =0.
\end{eqnarray*}

The Lie Brackets of the admitted Lie symmetries are presented in Table \ref%
{table1}, from which it is clear that the admitted Lie algebra is $\left\{
3A_{1}\oplus _{s}2A_{1}\right\} ~$in the Morozov-Mubarakzyanov
Classification Scheme \cite%
{Morozov58a,Mubarakzyanov63a,Mubarakzyanov63b,Mubarakzyanov63c}. The vector
fields $\left\{ X_{1},X_{2},X_{3}\right\} $ form the $E^{2}$ Lie algebra,
more specifically they are the isometries of the two-dimensional Euclidean
space, while the set of vectors $\left\{ X_{1},X_{2},X_{3},X_{5}\right\} $
form the Homothetic algebra for the two-dimensional Euclidean space.

\begin{table}[tbp] \centering%
\begin{tabular}{cccccc}
\hline\hline
$\left[ ~,~\right] $ & $X_{1}$ & $X_{2}$ & $X_{3}$ & $X_{4}$ & $X_{5}$ \\
\hline
$X_{1}$ & $0$ & \thinspace $0$ & $-X_{2}$ & $0$ & $X_{1}$ \\
$X_{2}$ & $0$ & $0$ & $X_{1}$ & $0$ & $X_{2}$ \\
$X_{3}$ & $X_{2}$ & $-X_{1}$ & $0$ & $0$ & $0$ \\
$X_{4}$ & $0$ & $0$ & $0$ & $0$ & $0$ \\
$X_{5}$ & $-X_{1}$ & $-X_{2}$ & $0$ & $0$ & $0$ \\ \hline\hline
\end{tabular}%
\caption{Lie Brackets for the admitted Lie point symmetries of equation
(\ref{eq.04})}\label{table1}%
\end{table}%

\subsection{Lie point symmetries for the time-dependent equation}

For the time-dependent equation (\ref{eq.03}) the admitted Lie point
symmetries are calculated to be%
\begin{eqnarray*}
Y_{1} &=&\partial _{x}~,~Y_{2}=\partial _{y}~,~Y_{3}=y\partial
_{x}-x\partial _{x}~,~Y_{4}=t\partial _{t}-2u\partial _{u}~,~~ \\
Y_{5} &=&5t\partial _{t}+x\partial _{x}+y\partial _{y}~,~Y_{6}=\partial
_{t},~~Y_{\infty }=\bar{\Psi}\left( x,y\right) \partial _{u}~\text{with }%
\Delta \bar{\Psi}=0.
\end{eqnarray*}

The Lie Brackets of the Lie point symmetries are presented in Table \ref%
{table2}; from which we infer that the admitted Lie algebra is $\left\{
2A_{1}\oplus _{s}A_{1}\right\} \oplus _{s}A_{2,1}$.

The table of Lie Brackets is

\begin{table}[tbp] \centering%
\begin{tabular}{ccccccc}
\hline\hline
$\left[ ~,~\right] $ & $Y_{1}$ & $Y_{2}$ & $Y_{3}$ & $Y_{4}$ & $Y_{5}$ & $%
Y_{6}$ \\ \hline
$Y_{1}$ & $0$ & \thinspace $0$ & $-Y_{2}$ & $0$ & $Y_{1}$ & $0$ \\
$Y_{2}$ & $0$ & $0$ & $Y_{1}$ & $0$ & $Y_{2}$ & $0$ \\
$Y_{3}$ & $Y_{2}$ & $-Y_{1}$ & $0$ & $0$ & $0$ & $0$ \\
$Y_{4}$ & $0$ & $0$ & $0$ & $0$ & $0$ & $-Y_{6}$ \\
$Y_{5}$ & $-Y_{1}$ & $-Y_{2}$ & $0$ & $0$ & $0$ & $-5Y_{6}$ \\
$Y_{6}$ & $0$ & $0$ & $0$ & $Y_{6}$ & $5Y_{6}$ & $0$ \\ \hline\hline
\end{tabular}%
\caption{Lie Brackets for the admitted Lie point symmetries of equation
(\ref{eq.03})}\label{table2}%
\end{table}%

The admitted Lie symmetries are again the isometries of the two-dimensional
Euclidean space, but the homothetic field is replaced by the scalar
symmetries $\left\{ Y_{4},Y_{5}\right\} $, while also the $Y_{6}$ symmetry
follows because the differential equation is invariant under time
translations.

\subsection{Lie point symmetries for the time-independent equation with a
linear source}

Before we proceed with the application of the Lie invariants, we determine
the Lie point symmetries of the following differential equation ,
\begin{equation}
\left( u_{xxxx}+2u_{xxyy}+u_{yyyy}\right) \left( u_{xx}+u_{yy}\right) -3
\left[ \left( u_{xx}+u_{yy}\right) _{,x}\right] ^{2}-3\left[ \left(
u_{xx}+u_{yy}\right) _{,y}\right] ^{2}+\lambda \sqrt{\left\vert
u_{,xx}+u_{,yy}\right\vert }u=0  \label{eq.05}
\end{equation}%
which is the time-indepedent equation (\ref{eq.03}) with a linear source $%
\lambda u$. In the following Section we see how this equation it can follow
from (\ref{eq.03}) for a specific value of the parameter $\lambda $.

The admitted Lie point symmetries are four and they are%
\begin{eqnarray*}
Z_{1} &=&\partial _{x}~,~Z_{2}=\partial _{y}~,~Z_{3}=y\partial
_{x}-x\partial _{y} \\
Z_{4} &=&x\partial _{x}+y\partial _{y}+10u\partial _{u}.
\end{eqnarray*}%
with Lie Brackets given in Table \ref{table3}. Easily we can infer that the
Lie symmetries form the Lie algebra $\left\{ 2A_{1}\oplus _{s}A_{1}\right\}
\oplus _{s}A_{1}$.

\begin{table}[tbp] \centering%
\begin{tabular}{ccccc}
\hline\hline
$\left[ ~,~\right] $ & $Z_{1}$ & $Z_{2}$ & $Z_{3}$ & $Z_{4}$ \\ \hline
$Z_{1}$ & $0$ & \thinspace $0$ & $-Z_{2}$ & $Z_{1}$ \\
$Z_{2}$ & $0$ & $0$ & $Z_{1}$ & $Z_{2}$ \\
$Z_{3}$ & $Z_{2}$ & $-Z_{1}$ & $0$ & $0$ \\
$Z_{4}$ & $-Z_{1}$ & $-Z_{2}$ & $0$ & $0$ \\ \hline\hline
\end{tabular}%
\caption{Lie Brackets for the admitted Lie point symmetries of equation
(\ref{eq.05})}\label{table3}%
\end{table}%

We proceed with the application of the Lie symmetry vectors for equation (%
\ref{eq.03}).

\section{Application of Lie symmetries}

\label{sec3}

The nonlinear PDE is an $1+2$ fourth-order equation. We continue our
analysis by eliminating the time-derivative. For that, we start by
performing reductions with the use of the vector fields (a) $Y_{6}$ and (b)~$%
Y_{4}$ .

\subsection{Reduction with $\partial _{t}$}

The application of the autonomous symmetry $Y_{6}$ to equation (\ref{eq.03})
provides that the solution, $u$, is static, i.e., $u\left( t,x,y\right)
=u\left( x,y\right) $. Consequently the resulting equation is that of
expression (\ref{eq.04}). Now we can continue with the reduction process by
using the Lie symmetries $X_{\mathbf{I}}.$

\subsubsection{Travelling-wave solution}

Consider the vector field $X_{1}+\alpha X_{2}$. Then the resulting Lie
invariants are $\zeta =y-\alpha x$,~$u=w\left( \zeta \right) $. We consider $%
\zeta $ to be the new indepedent variable and $w$ the new dependent
variable. Thus equation (\ref{eq.04}) is simplified to the following
fourth-order ordinary differential equation (ODE)%
\begin{equation}
\left( a^{2}+1\right) \left( 2\frac{d^{4}w}{d\zeta ^{4}}\frac{d^{2}w}{d\zeta
^{2}}-3\left( \frac{d^{3}w}{d\zeta ^{3}}\right) ^{2}\right) =0.
\label{eq.06}
\end{equation}

For this equation easily we can calculate the admitted Lie point symmetries
which are%
\begin{equation*}
\bar{X}_{1}=\partial _{w}~,~\bar{X}_{2}=\partial _{\zeta }~,~\bar{X}%
_{3}=w\partial _{w}~,\text{ }X_{4}=\zeta \partial _{\zeta }\text{ and }%
X_{5}=\zeta \partial _{w}
\end{equation*}%
which form the Lie algebra $\left\{ 2A_{1}\oplus _{s}A_{1}\right\} \oplus
_{s}A_{1,1}$.

While we could continue with the application of Lie point symmetries to
equation (\ref{eq.06}), it can easily be integrated and we get%
\begin{equation}
w\left( \zeta \right) =w_{1}\ln \left( \zeta -\zeta _{1}\right) +w_{2}\left(
\zeta -\zeta _{2}\right),
\end{equation}%
where $w_{1,2}$ and $\zeta _{1,2}$ are four constants of integration.

Because the parameter $\alpha $ is not involved in the resulting ODE (\ref%
{eq.06}) the solution holds also for reduction process of (\ref{eq.04}) with
the vector field $X_{1}$ or $X_{2}.$ Moreover, equation (\ref{eq.06}) can be
written easily as the second-order ordinary differential equation%
\begin{equation}
2W\frac{d^{2}W}{d\zeta ^{2}}-3\left( \frac{dW}{d\zeta }\right) ^{2}=0,
\label{eq.06b}
\end{equation}%
where $W\left( \zeta \right) =\frac{d^{2}w}{d\zeta }$. We can easily see
that it is maximally symmetric and admits eight Lie point symmetries, which
form the $sl\left( 3,R\right) $ Lie algebra. Equation (\ref{eq.06b}) can be
easily linearised after the transformation $W\left( \zeta \right) =\left(
\bar{W}\left( \zeta \right) \right) ^{-\frac{1}{2}}$~\cite{line1,line2,line3}%
..

\subsubsection{Radial solution}

Consider now the application of the Lie symmetry vector $X_{3}$, i.e. of the
rotational symmetry. The invariant functions are calculated to be $%
r=x^{2}+y^{2}$ and $u=h\left( r\right) $. The resulting ODE is
\begin{equation}
-2\frac{d^{2}h}{dR^{2}}\left( \frac{d^{4}h}{dR^{4}}+\frac{d^{3}h}{dR^{3}}%
\right) +\left( \frac{d^{2}h}{dR^{2}}\right) ^{2}+3\left( \frac{d^{3}h}{%
dR^{3}}\right) ^{2}=0  \label{eq.07}
\end{equation}%
after we applied the transformation $r=\exp \left( R\right) $ in order to
simplify the reduced equation.

Equation (\ref{eq.07}) admits the four-dimensional Lie algebra, $%
2A_{1}\oplus _{s}A_{1,1}$, which comprises the following symmetry vectors%
\begin{equation*}
X_{1}^{\prime }=\partial _{R}~,~X_{1}^{\prime }=\partial
_{h}~,~X_{3}^{\prime }=h\partial _{h}~\text{\ and }X_{4}=R\partial _{h}.
\end{equation*}

The fourth-order ODE can be easily reduced to the second-order ODE%
\begin{equation}
-2\rho \left( \frac{d^{2}\rho }{dR^{2}}+\frac{d\rho }{dR}\right) +\rho
^{2}+3\left( \frac{d\rho }{dR}\right) ^{2}=0  \label{eq.08}
\end{equation}%
by applying the change of variable $\rho =\frac{d^{2}h}{dR^{2}}$.

Surprisingly, equation (\ref{eq.08}) is maximally symmetric, that is, it
admits as Lie point symmetries the elements of the $sl\left( 3,R\right) $
Lie algebra in the representation%
\begin{eqnarray*}
&&\partial _{R}~,~\rho \partial _{\rho }~,~e^{-\frac{R}{2}}\rho ^{\frac{3}{2}%
}\partial _{\rho }~,~Re^{-\frac{R}{2}}\rho ^{\frac{3}{2}}\partial _{\rho
}~,~R\partial _{R}+R\rho \partial _{\rho }~, \\
&&\frac{e^{\frac{\rho }{2}}}{\sqrt{\rho }}\left( \partial _{R}+\rho \partial
_{\rho }\right) ~,~\frac{e^{\frac{\rho }{2}}}{\sqrt{\rho }}\left( R\partial
_{R}+\rho \left( R-2\right) \partial _{\rho }\right) ~,~R^{2}\partial
_{R}+R\left( \rho R-2\rho \right) \partial _{\rho }.
\end{eqnarray*}%
The generic solution of (\ref{eq.08}) is given to be
\begin{equation}
\rho \left( R\right) =\rho _{0}e^{R}\left( \left( R-R_{0}\right) \right)
^{-2},  \label{eq.09}
\end{equation}%
where $\rho _{0},R_{0}$ are constants of integration.

\subsubsection{Scaling solution}

From the scaling symmetry $X_{5}$ we determine the invariants $\theta
=\arctan \left( \frac{y}{x}\right) $ and $u=g\left( \theta \right) $. The
reduced ODE is%
\begin{equation}
2\left( \frac{d^{2}g}{d\theta ^{2}}\right) \left( \frac{d^{4}g}{d\theta ^{4}}%
\right) -3\left( \frac{d^{3}g}{d\theta ^{3}}\right) ^{2}-4\left( \frac{d^{2}g%
}{d\theta ^{2}}\right) ^{2}=0,  \label{eq.10}
\end{equation}%
which is invariant under the Lie symmetry vectors%
\begin{equation*}
X_{1}^{\ast }=\partial _{\theta }~,~X_{2}^{\ast }=\partial
_{g}~,~X_{3}^{\ast }=g\partial _{g}~\text{and }X_{4}^{\ast }=\theta \partial
_{g}\text{.}
\end{equation*}

The generic solution of equation (\ref{eq.10}) easily can be determined. It
is%
\begin{equation}
g\left( \theta \right) =g_{1}\ln \left( \cos \left( \theta -\theta
_{1}\right) \right) +g_{2}\left( \theta -\theta _{2}\right)  \label{eq.11}
\end{equation}%
in which $g_{0,1}$ and $\theta _{0,1}$ are four constants of integration.

However, what is important to mention is that equation (\ref{eq.10}) can be
written again as a second-order ODE by applying the change of variable $%
G=\left( \frac{d^{2}g}{d\theta ^{2}}\right) $. The resulting second-order
ODE is maximally symmetric and under the $G=\bar{G}^{-2}$ takes the simple
form of the harmonic oscillator%
\begin{equation}
\frac{d^{2}\bar{G}}{d\theta ^{2}}+\bar{G}=0.  \label{eq.12}
\end{equation}

\subsection{Reduction with $t\partial _{t}-2u\partial _{u}$}

We continue our analysis by applying the invariants to equation (\ref{eq.03}%
) of the symmetry vector $Y_{4}$. The invariant functions are $x,y$ and $u=%
\frac{u\left( x,y\right) }{t^{2}}$. The resulting PDE is equation (\ref%
{eq.05}) for the specific value $\lambda =-2$.

\subsubsection{Travelling-wave solution}

Reduction with the use of the symmetry vectors $Z_{1}+\alpha Z_{2}$ provides
the fourth-order ODE%
\begin{equation}
2\left( \frac{d^{2}w}{d\zeta ^{2}}\right) \left( \frac{d^{4}w}{d\zeta ^{4}}%
\right) -3\left( \frac{d^{3}w}{d\zeta ^{3}}\right) ^{2}-2\sqrt{\left( \alpha
^{2}+1\right) ^{-5}\frac{d^{2}w}{d\zeta ^{2}}}w=0  \label{eq.13}
\end{equation}%
in which $\zeta =y-\alpha x$ and $w=w\left( \zeta \right) $.

Equation (\ref{eq.13}) is invariant only under the two-dimensional Lie
algebra $A_{1,1}$ with as elements the vector fields
\begin{equation*}
\bar{Z}_{1}=\partial _{\zeta }~,~\bar{Z}_{2}=\zeta \partial _{\zeta
}+10w\partial _{w}.
\end{equation*}%
A special solution of (\ref{eq.13}) can be found easily by the use of the
symmetry vector $\bar{Z}_{2}$ and it is $w\left( \zeta \right) =w_{0}\zeta
^{10}$ for $w=w\left( \alpha \right) $.

The application of the two symmetries $\bar{Z}_{1},~\bar{Z}_{2}$ to (\ref%
{eq.13}) leads to a second-order equation with no Lie point symmetries.

\subsubsection{Radial solution}

From the vector field $Z_{3}$ we get the invariants $R=\ln \left(
x^{2}+y^{2}\right) $ and $u=h\left( r\right) $. The reduced ODE is
\begin{equation}
-2\frac{d^{2}h}{dR^{2}}\left( \frac{d^{4}h}{dR^{4}}+\frac{d^{3}h}{dR^{3}}%
\right) +\left( \frac{d^{2}h}{dR^{2}}\right) ^{2}+3\left( \frac{d^{3}h}{%
dR^{3}}\right) ^{2}-\frac{1}{16}\sqrt{\left( \frac{d^{2}h}{dR^{2}}\right) }%
he^{\frac{5}{2}R}=0,  \label{eq.14}
\end{equation}%
which admits only one Lie point symmetry, namely
\begin{equation*}
Z^{\prime }=\partial _{R}+5h\partial _{h}.
\end{equation*}

Application of this to (\ref{eq.14}) leads to a third-order ODE with no Lie
point symmetries. A special solution of (\ref{eq.14}) can be found by
applying the invariant of $Z^{\prime }$. It is $h\left( R\right)
=h_{0}e^{5R} $, where $h_{0}$ is an imaginary number.

\subsubsection{Scaling solution}

The application of the scaling symmetry, $Z_{4}$, leads to a fourth-order
ODE with only one symmetry, the vector field $Z^{\ast }=Z_{3}$ the
application of which reduces the equation to a third-order equation without
Lie point symmetries. For convenience of the presentation we do not write
the fourth-order or the third-order equations.

\section{Conclusions}

\label{sec4}

In this work, we have focused on the application of Lie's theory for the
determination of invariant one-parameter point transformations for a
fourth-order $1+2$ evolution equation which has been proposed for the study
of noise reduction in image-processing theory. The model of our
consideration is a higher-order generalization of the model proposed by
Rudin et al \cite{ref1} and it has been proposed later by You and Kaveh \cite%
{you}.

We have performed the symmetry classification for the following kind of
solutions: (i) time-dependent solution, (ii) stationary solution with no
source and (iii) stationary solution with linear source. The later
classification has been useful in order to continue the reduction process on
the $1+2$ evolution equation to a fourth-order ODE.

The line point symmetries have been applied in order to perform
second-reductions that they provide similarity solutions which belong to
family of: travelling-wave solutions, radial solutions and scaling
solutions. Surprisingly, in the case of static solutions all the
second-reductions reduce to fourth-order ODEs which can be written as linear
second-order ODEs. That it is not the case of nonstatic solutions for which
we were able to reduce the fourth-order ODEs to third-order ODEs with the
application of Lie symmetries. However, some specific closed-form solutions
were determined.

It is important to mention the existence of the two symmetries $u\partial
_{u}$ and $\Psi \left( x,y\right) \partial _{u}$ which indicates linearity,
even if the PDE (\ref{eq.03}) is always nonlinear. These two symmetries are
related with the solution of the two-dimensional Laplace equation which is
the common factor in (\ref{eq.03}). However, these solutions are not
acceptable for our consideration. For that reason we have not applied these
two symmetries in the reduction process. On the other hand, from the
similarity solutions we determined with the use of the Lie symmetries we
show that new solutions exist.

The results of this work indicates that the theory of symmetries of
differential equations can be play an important role for the study of the
nonlinear PDEs in the image-processing theory. Therefore, conservation laws
can be also determined by using Noether's theorem or by using other
approaches. Such a work is still in progress and will be published elsewhere.

\bigskip

\textbf{Acknowledgements}

\textit{PGLL Thanks the Durban University of Technology, the University of
KwaZulu-Natal and the National Research Foundation of South Africa for
support. The authors thank Suranaree University of Technology, Sergey
Meleshko and Eckart Schulz for the the hospitality provided while the bulk
of this work undertaken.}

\end{document}